\documentclass[aps,prb,reprint,superscriptaddress]{revtex4-2}
\usepackage{physics}
\usepackage{lmodern}
\usepackage{graphicx}%
\usepackage{amsmath}%
\usepackage{amsfonts}%
\usepackage{amssymb}
\usepackage{bm}
\usepackage{xcolor}
\usepackage{hyperref}
\usepackage{ulem}

\def\d{{\partial}}
\def\s{{\sigma}}

\def\k{{ {\bm k} }}

\def\0{{ {\bm 0} }}

\def\ve{{\varepsilon}}

\def\rR{{ {\rm R} }}
\def\rA{{ {\rm A} }}

\def\cV{{ {\cal V} }}

\def\ang{{ \mathrm{\mathring{A}} }}

\allowdisplaybreaks[4]

\begin{document} 
\title{Nonlinear charge and thermal transport properties induced by orbital magnetic moment in chiral crystal cobalt monosilicide}
\author{Kazuki Nakazawa} 
\author{Terufumi Yamaguchi} 
\affiliation{RIKEN Center for Emergent Matter Science, 2-1 Hirosawa, Wako, Saitama 351-0198, Japan}
\author{Ai Yamakage} 
\affiliation{Department of Physics, Nagoya University, Nagoya 464-8602, Japan}
\date{\today}

\begin{abstract}
The existence of exotic singularities in momentum space, such as spin-1 excitations and Rarita-Schwinger-Weyl (RSW) fermions, has been discussed so far to explore unique phenomena in the nonmagnetic B20-type compounds. Meanwhile, the Nonlinear Thermo-Electric (NCTE) charge and thermal Hall effect, a response proportional to the cross product of the electric field and temperature gradient, is expected in this chiral material, yet remains unexplored in B20-type compounds. Here, based on \textit{ab initio} calculations and symmetry analysis, we quantitatively analyze the NCTE charge and thermal Hall effects in cobalt monosilicide, obtaining experimentally measurable values of NCTE charge and thermal Hall current along [111] direction, which is not expected for second-order current responses to the DC electric field. Furthermore, we demonstrate that these significant responses are enhanced around RSW fermions and spin-1 excitations. Additionally, we clarify that the NCTE Hall effect is solely governed by orbital magnetic moments due to the cancellation of Berry curvature contributions in cubic chiral crystals.  
\end{abstract}

\maketitle

%%%%%%%%%%%%%%%%%%
\section{Introduction}
\label{sec:intro}
%%%%%%%%%%%%%%%%%%

Counterparts of the elementary particles described by relativistic quantum mechanics often appear in the band structure of a crystal. For example, Weyl semimetals are known to have monopoles characterized by chirality in the momentum space of three-dimensional (bulk) materials, which induce peculiar surface states and emergent electromagnetic responses~\cite{AMV2018,NN1983,Murakami2007}. Recently, the diversity of such singularities in the band structure has been recognized, and its systematic classification by crystal symmetry has been conducted~\cite{Bradlyn2016}.

\begin{figure}
\includegraphics[width=85mm]{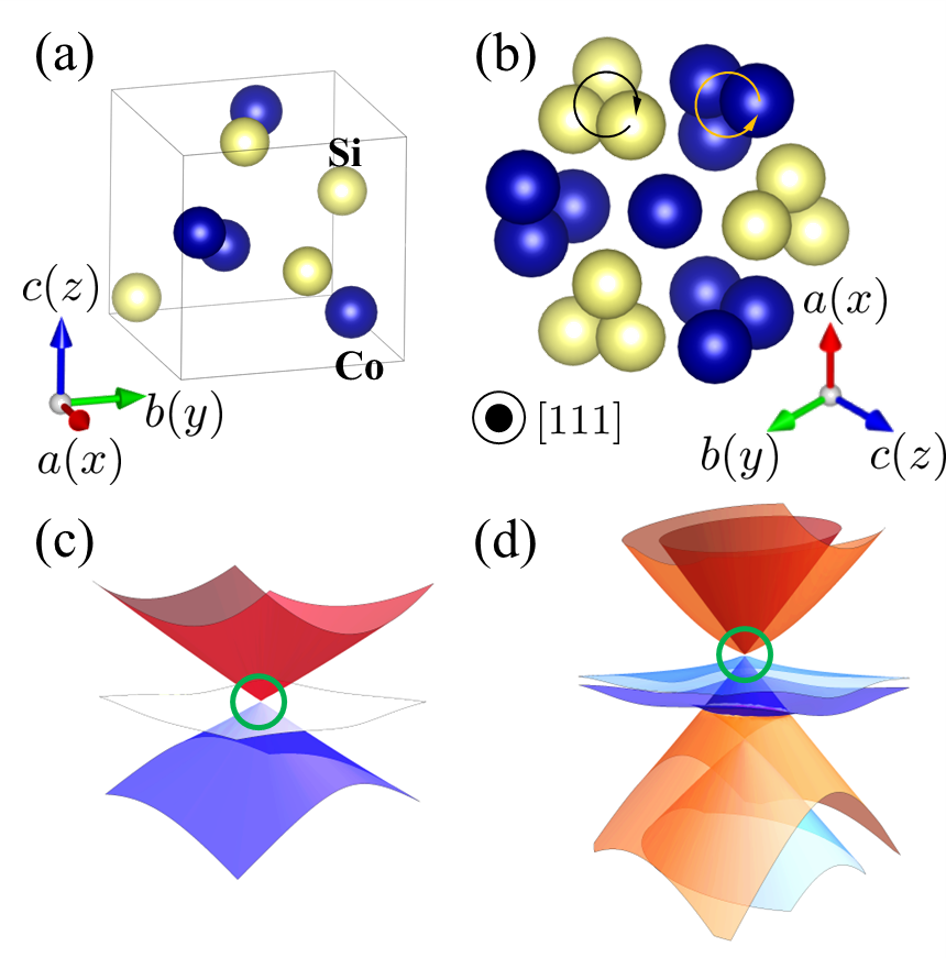}
\caption{(a) Lattice structure of the cobalt monosilicide. Black lines represent a primitive unit cell. (b) View from [111] direction. The crystal structures are visualized by VESTA~\cite{Momma2011}. [(c) and (d)] Peculiar excitation spectra reside in the band structure of CoSi indicated by green circles: (c) Spin-1 excitation and (d) Spin-3/2 Rarita-Schwinger-Weyl fermion. 
}
\label{fig:1}
\end{figure}

B20-type compounds belong to space group $P$2$_1$3 (No. 198) and have a chiral crystal structure that does not host an inversion center or mirror planes [Figs.~\ref{fig:1}(a) and (b)]. The chirality of the crystal structure leads to an antisymmetric magnetic interaction, i.e., Dzyaloshinskii-Moriya interaction, which gives rise to various magnetic structures such as helical spirals~\cite{IA1984,Grigoriev2006}, skyrmions~\cite{Muhlbauer2009,NT2013}, and hedgehogs~\cite{Fujishiro2019,FKT2020} in magnets such as MnSi, MnGe, and FeGe. On the other hand, CoSi, CoGe, and RhSi, which are nonmagnetic, have been pointed out to exhibit rare emergent multi-fold chiral fermions such as spin-1 excitation [Fig.~\ref{fig:1}(c)] and/or spin-3/2 Rarita-Schwinger-Weyl (RSW) fermions [Fig.~\ref{fig:1}(d)] at the vicinity of the Fermi level~\cite{TZZ2017}, and their surface states~\cite{Takane2019}, thermo-electric properties~\cite{SZPM2009}, spin transport~\cite{Tang2021}, and nonlinear transport~\cite{Ni2020} have been investigated. Moreover, the orbital magnetic moment distribution in momentum space and the orbital Hall effect have been theoretically pointed out recently~\cite{Yang2023}. 

\begin{figure*}
\hspace*{-3mm}
\includegraphics[width=180mm]{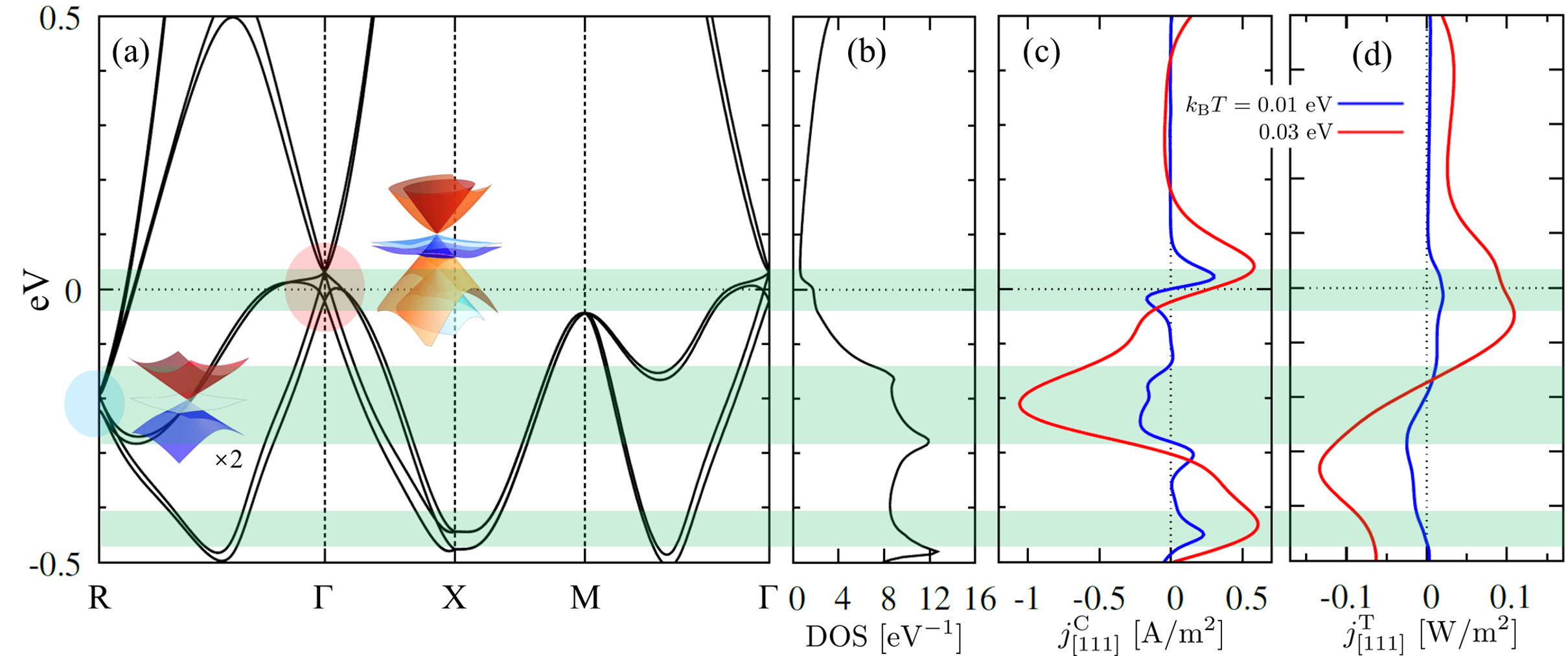}
\caption{(a) Band structure of CoSi. The RSW fermion at the $\Gamma$ point and spin-1 excitation at the R point are marked by red and blue shades, respectively. (b) Density of states of CoSi. [(c) and (d)] Chemical potential dependence of (c) the NCTE charge Hall current and (d) the NCTE thermal Hall current along the [111] direction calculated for $k_{\rm B}T = 0.01$~eV and 0.03~eV. 
}
\label{fig:2}
\end{figure*}

Besides, a second-order current response to the external field is expected in crystals that do not have inversion symmetry~\cite{SG1993,SS2000,GYN2014,Yokouchi2017,Morimoto2018,MP,MN,IN2020,Okumura2021-vz,DLX2021,DWSLX}. In addition to nonlinear responses originating from band asymmetry, which appear when both time-reversal and spatial inversion symmetries are broken~\cite{Yokouchi2017,Morimoto2018,IN2020,Okumura2021-vz,MN}, contributions from higher-order band geometries have recently been extensively studied. For instance, the nonlinear Hall effect due to Berry curvature dipoles appears even with time-reversal symmetry~\cite{SF2015,Ma2019}, and nonlinear transport due to the quantum metric~\cite{GYN2014,NWang2023} has gained recognition in recent years. Responses to higher-order of temperature gradients have also been investigated to predict the nonlinear spin Seebeck effect~\cite{TSM2018} and nonlinear thermal transport~\cite{Nakai2019,NKM2022,Arisawa2024}. Additionally, the response to the product of the electric field and temperature gradient has also been explored. Very recently, we have microscopically formulated the Nonlinear Chiral Thermo-Electric (NCTE) Hall effect~\cite{YNY2023}, a charge current response to the cross product of the electric field and temperature gradient~\cite{Nakai2019,Hidaka2018,Toshio2020,YNY2023}. We have shown that not only the Berry curvature dipole but also the orbital magnetic moment makes an important contribution to the NCTE Hall effect, and we indeed demonstrated its significance in chiral tellurium~\cite{NYY2024}. The NCTE Hall current changes its sign depending on crystal chirality, and suggests potential applications in heat flow sensors where sensitivity can be controlled by an applied electric field. Moreover, the \lq\lq NCTE thermal Hall effect," which is the thermal current response to the cross product of the electric field and the temperature gradient, is also expected. Combined with the large orbital magnetic moment in CoSi, it represents an ideal model case to study the charge and thermal transport arising from orbital magnetism, although this has not yet been clarified.

In this paper, we investigate the NCTE charge and thermal Hall effects under the actual band dispersion of CoSi, based on \textit{ab initio} calculations. We successfully reproduce the overall band structure, which includes a four-fold degenerate RSW fermion at the $\Gamma$ point and a six-fold degenerate double spin-1 excitation at the R point. The three-fold rotational symmetry around the [111] axis leads to the absence of the Berry curvature dipole term, which highlights the importance of the contribution from the orbital magnetic moment. The NCTE charge and thermal Hall currents increase around the chemical potential where the RSW fermion and double spin-1 excitation reside. For comparison with the NCTE Hall effect, we also calculate the second-order response to the DC electric field, which does not arise in the [111] direction. This behavior is also observed in chiral tellurium~\cite{NYY2024} and contrasts with the NCTE Hall effect. We show that the NCTE charge and thermal Hall effects can be a very useful transport measurement for detecting the orbital magnetic moments arising from the multi-fold chiral fermion in chiral cubic crystal.

\begin{table*}[t]
\centering
\newlength{\height} 
\setlength{\height}{3mm}
\begin{center}
%\hspace*{-14mm}
\fontsize{13pt}{16pt}\selectfont
\begin{tabular}{ccccll}
\hline\hline
 $T$ & $E$ & $4C_3$ & $3C_2$ & Linear & Quadratic
\\
\hline
 $A$ & 1 & 1 & 1 & & $XX+YY+ZZ$
\\
 ${}^1E$ & 1 & $e^{i 2\pi/3}$ & 1 & & $2ZZ-XX-YY-i \sqrt{3} (XX-YY)$
\\
 ${}^2E$ & 1 & $e^{-i 2\pi/3}$ & 1 & & $2ZZ-XX-YY +i \sqrt{3} (XX-YY)$
\\
 $T$ & 3 & 0 & $-1$ & $(X, Y, Z)$ & $(YZ, ZX, XY),  (ZY, XZ, YX)$
\\
\hline\hline
\label{tab:T}
\end{tabular}
\caption{Character table of the point group $T$. $A$, ${}^1 E$, ${}^2 E$, and $T$ represent irreducible representations, and $X$, $Y$, and $Z$ are basis functions corresponding to the coordinate $x$, $y$, and $z$. }
\end{center}
\end{table*}

%%%%%%%%%%%%%%%%%%
\section{band structure and orbital magnetic moment}
\label{sec:abinitio}
%%%%%%%%%%%%%%%%%%

The band structure is obtained from a calculation based on the density functional theory (DFT) using OpenMX code~\cite{OpenMX,Ozaki2003}. The framework of generalized gradient approximation (GGA) proposed by Perdew-Burke-Ernzerhof~\cite{PBE1996} is used for the exchange-correlation functional and norm-conserving and fully momentum-dependent pseudopotentials is chosen to incorporate the effect of the spin-orbit coupling. The wave functions are expanded using linear combinations of pseudoatomic orbitals. The basis set for pseudoatomic orbitals is specified as Co6.0H-s3p2d1 and Si7.0-s2p2d1. We use the lattice constants of $a=b=c=4.454$~$\ang$. We set the cutoff energy which specifies the fast Fourier transform grid to 1200 Ry and sampled the Brillouin zone with $16^3$ $k$ points. The self-consistent field calculation converged to the paramagnetic state, which matches to the previous observations~\cite{TZZ2017,Takane2019}. 

From the Bloch states obtained in the DFT calculations mentioned above, a Wannier basis set is provided using the OpenMX code~\cite{OpenMX,WOT2009} consisting of $d$ orbitals localized at each Co sites and $s$ and $p$ orbitals localized at the Si sites, for a 72-orbital model including spin degrees of freedom. These sets are based on DFT Bloch bands in the energy range from $-14$~eV to +8~eV and almost perfectly reproduce the bands in the range from $-14$~eV to +1.5~eV which is the inner window we set in the calculation.

\begin{figure}[t]
\includegraphics[width=85mm]{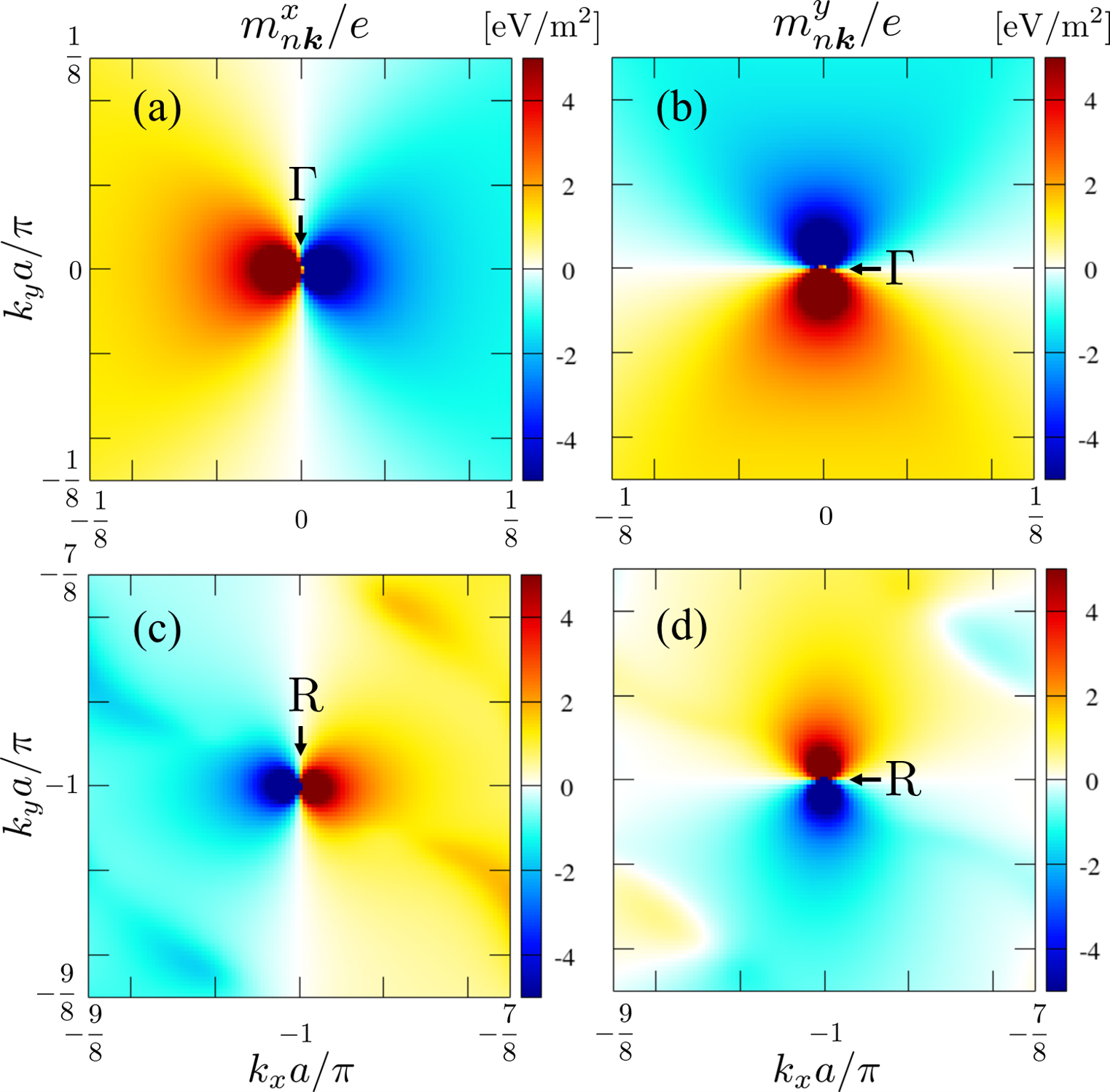}
\caption{Momentum space distributions of %[(a) and (b)] %Berry curvature ${\bm \Omega}_{n\k}$ and [(c) and (d)] 
orbital angular momentum ${\bm m}_{n\k}$ of the %top most
highest energy bands 
%\textcolor{magenta}{[the lowest conduction band?]} \cyan{[No. The plots (a,b) and (c,d) are for different bands.]} 
comprising [(a) and (b)] RSW fermion at $\Gamma$ point and [(c) and (d)] double spin-1 excitation at R point. The former two and latter two panels are for the $k_z = 0$ cut %of the RSW fermion 
and $k_z = \pi/a$ cut of %the double spin-1 excitation, 
each band respectively, which are calculated using Eq.~\eqref{eq:OM}. 
}
\label{fig:3}
\end{figure}

Figure~\ref{fig:2}(a) shows the band structure obtained from DFT calculations, which reproduces well those reported in previous study~\cite{TZZ2017}. A notable feature of this band structure is the appearance of RSW fermion near the Fermi level at the $\Gamma$ point and double spin-1 excitations around $\mu= -0.2$~eV at the R point. In the absence of spin-orbit coupling, the $\Gamma$ and R points host spin-1 excitations and double Weyl fermions, respectively, with spin degrees of freedom leading to 6-fold and 8-fold degenerate points for each. The former splits into a 4-fold degenerate states (RSW fermion) and a 2-fold degenerate states, while the latter splits into a 6-fold degenerate states (double spin-1 excitation) and a 2-fold degenerate states due to the spin-orbit coupling. The RSW fermions and double spin-1 excitations possess monopole charges of $\pm 4$ and $\mp 4$, respectively, accompanied by the divergence of the Berry curvature at each singularity~\cite{TZZ2017}. Orbital magnetic moment has the same symmetry as the Berry curvature and also exhibit similar momentum-space properties. The Berry curvature ${\bm \Omega}_{n\k}$ and the orbital magnetic moment ${\bm m}_{n\k}$ are calculated using
\begin{align}
{\bm \Omega}_{n\k} 
&= {\rm Im} [ \nabla_\k \times \langle n (\k) | \nabla_\k  m (\k) \rangle ] \nonumber
\\
&= i \sum_{m \neq n} \frac{ \langle n (\k) | \hat{\bm \cV}_\k | m (\k) \rangle \times \langle m (\k) | \hat{\bm \cV}_\k | n (\k) \rangle }{ (\ve_{n\k} - \ve_{m\k} )^2 } , \label{eq:BC}
\\
{\bm m}_{n\k} 
&= \frac{e}{2} {\rm Im} [ \langle \nabla_\k n (\k) | \times \{ \hat{H}_\k - \ve_{n \k} \} | \nabla_\k  m (\k) \rangle ] \nonumber
\\
&= \frac{ie}{2} \sum_{m \neq n} \frac{ \langle n(\k) | \hat{\bm \cV}_\k | m (\k) \rangle \times \langle m (\k) |  \hat{\bm \cV}_\k | n (\k) \rangle }{ (\ve_{n\k} - \ve_{m\k} ) } . \label{eq:OM}
\end{align} 
Here we imply an elementary charge $e$, an eigenenergy $\ve_{n\k}$ and an eigenvector $| n (\k) \rangle$ of the Hamiltonian $\hat{H}_\k$, and the velocity operator $\hat{\bm \cV}_\k = \nabla_\k \hat{H}_\k$. Figure~\ref{fig:3} displays momentum-space plots of each component of the orbital magnetic moment ${\bm m}_{n\k}$ at $\Gamma$ and R points, calculated using the Wannier model, revealing dipole-like distributions. Hereafter, let us define $x$, $y$, and $z$ axes identical to $a$, $b$, and $c$ axes, respectively. The presence of such peculiar distributions of the orbital magnetic moment strongly suggest the existence of NCTE charge and thermal Hall effects, which we discuss in detail below through symmetry analysis and quantitative evaluations.

%%%%%%%%%%%%%%%%%%
\section{Nonlinear transport properties}
\label{sec:transport}
%%%%%%%%%%%%%%%%%%

\begin{figure}
\includegraphics[width=85mm]{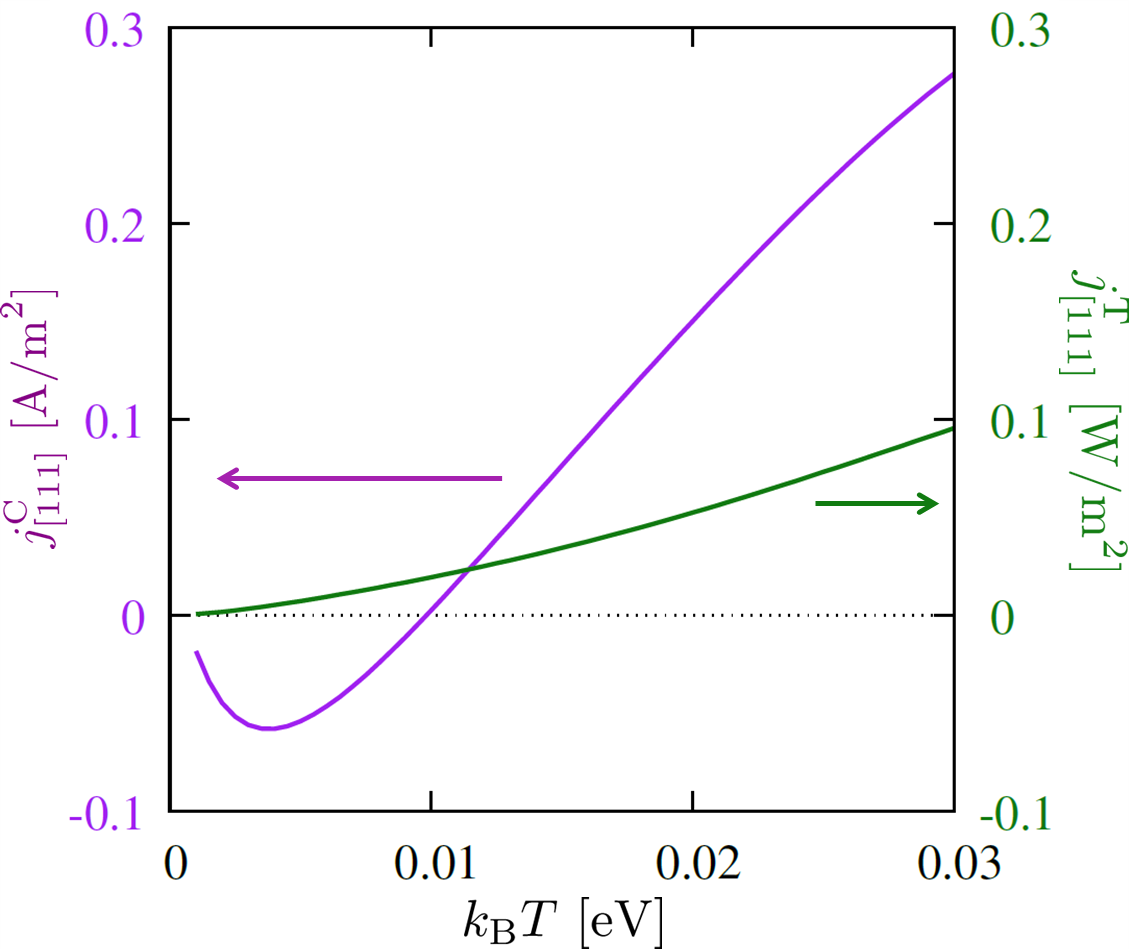}
\caption{Temperature dependence of the NCTE charge and thermal Hall current $j_{[111]}^{\rm C}$ and $j_{[111]}^{\rm T}$, respectively. 
}
\label{fig:4}
\end{figure}

\begin{figure*}
\includegraphics[width=175mm]{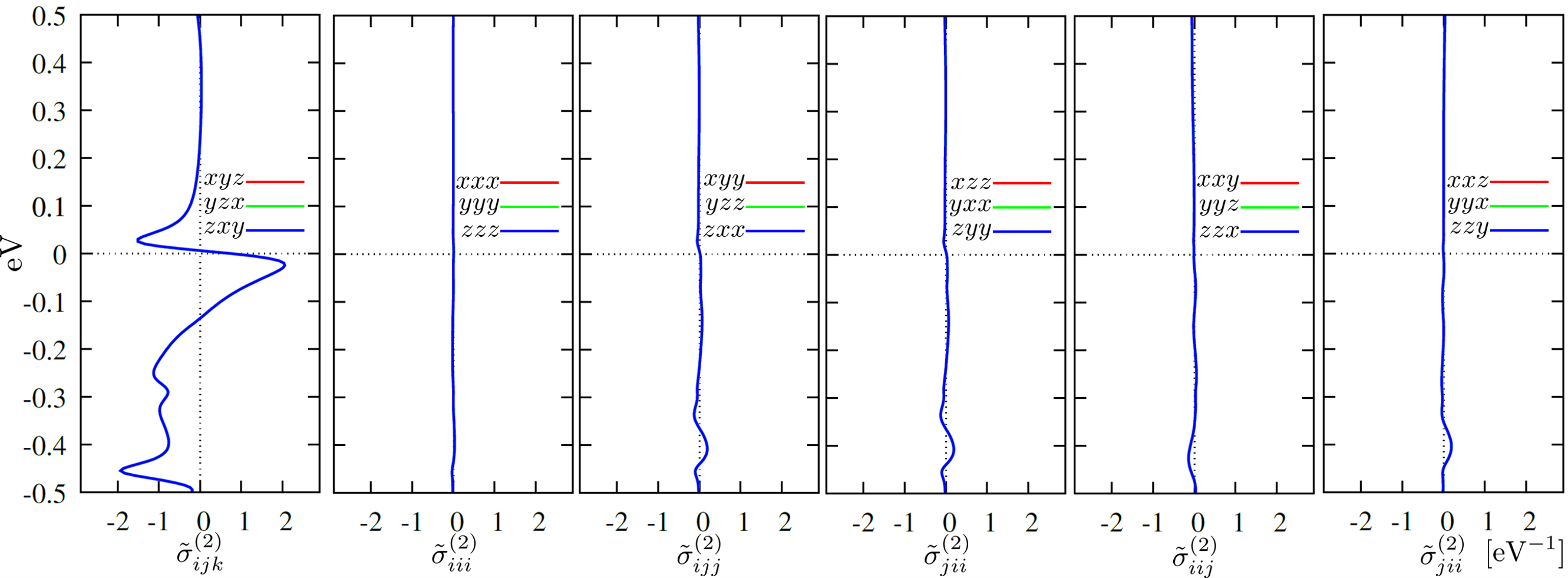}
\caption{Chemical potential dependence of the second-order nonlinear conductivity which is normalized as $\tilde{\sigma}_{ijk}^{(2)} \equiv \sigma_{ijk}^{(2)}/(e^3/h)$. The red, green, and blue lines correspond to the current along the $x$, $y$, and $z$ directions. 
}
\label{fig:5}
\end{figure*}

%%%%%%%%%%%%%%%%%%
\subsection{Microscopic formula}
\label{sec:formula}
%%%%%%%%%%%%%%%%%%

The microscopic calculation revealed that the NCTE charge Hall current $\langle j_z^{\rm C} \rangle = \sigma_z^{\rm C} \{ {\bm E} \times (- \nabla T/T ) \}_z$ is dominated by following two terms~\cite{YNY2023};  
\begin{align} 
&\sigma_z^{\rm C} \simeq \sigma^{\mathrm{BCC}}_z + \sigma^{\mathrm{OMC}}_z , \label{eq:NCTE_TOT} \\
&\sigma^{\mathrm{BCC}}_z= e^{2} \tau \sum_{n,\bm{k}}
F_{\rm C}(\ve_{n\k})
\left\{ 	\Omega'_z - \frac{1}{2} \left( 
	\Omega'_x + \Omega'_y \right) \right\} ,
\label{eq:NCTE_yC}
\\
&\sigma^{\mathrm{OMC}}_z = -\frac{1}{2} e \tau \sum_{n,\bm{k}}
F_{\rm C}(\ve_{n\k})
\bm{\nabla}_{\bm{k}} \cdot \bm{m}^{\perp}_{n\bm{k}} ,
\label{eq:NCTE_OM}
\end{align}
where ${\bm E}$ is the DC electric field, $\tau$ is an electron lifetime, $F_{\rm C}(\ve) = (\ve - \mu) (-\frac{\partial f}{\partial \ve})$ with the temperature $T$, chemical potential $\mu$, and the Fermi-Dirac distribution function $f$, $\Omega'_i \equiv v_{n{\bm k}}^i \Omega_{n\bm k}^i =  (\partial_{k_i} \varepsilon_{n{\bm k}} ) \Omega_{n\bm k}^i$ with group velocity ${\bm v}_{n\k} = \nabla_\k \ve_{n\k}$, and ${\bm m}_{n\k}^\perp = (m_{n\k}^x, m_{n\k}^y, 0)$ is the orbital magnetic moment which only contains in-plane components. We can similarly give the expression of the NCTE thermal Hall current and conductivity $\langle j_z^{\rm T} \rangle = \sigma_z^{\rm T} \{ {\bm E} \times (- \nabla T/T ) \}_z$,   
\begin{align} 
&\sigma_z^{\rm T} \simeq \sigma^{\mathrm{BCT}}_z + \sigma^{\mathrm{OMT}}_z \label{eq:NCTE_TOTT}, \\
&\sigma^{\mathrm{BCT}}_z= -e \tau \sum_{n,\bm{k}}
F_{\rm T}(\ve_{n\k})
\left\{ \Omega'_z - \frac{1}{2} \left( 
	\Omega'_x + \Omega'_y \right) \right\},
\label{eq:NCTE_yCT}
\\
&\sigma^{\mathrm{OMT}}_z = \frac{1}{2} \tau \sum_{n,\bm{k}}
F_{\rm T}(\ve_{n\k}) \bm{\nabla}_{\bm{k}} \cdot \bm{m}^{\perp}_{n\bm{k}} ,
\label{eq:NCTE_OMT}
\end{align}
where $F_{\rm T}(\ve) = (\ve - \mu)^2 (-\frac{\partial f}{\partial \ve})$. Hereafter we call the terms $\s_z^{\rm BCC}$ and $\s_z^{\rm BCT}$ as Berry curvature dipole terms, and $\s_z^{\rm OMC}$ and $\s_z^{\rm OMT}$ as orbital magnetic moment terms. We used the Wannier model for the actual calculations. 

The second-order current response to the DC electric field $\langle j_i^{(2)} \rangle = \sigma_{ijk}^{(2)} E_j E_k$ is also calculated in the same Wannier model. The second-order DC nonlinear conductivity $\sigma_{ijk}^{(2)}$ is given by~\cite{MP,MN}
\begin{align}
&\sigma_{ijk}^{(2)} %&
\simeq
\frac{2e^3}{V} \int \frac{d\ve}{2\pi} \left(-\frac{\partial f}{\partial \ve}\right)  
\nonumber \\
&\quad \times {\rm Im}   \sum_\k 
%\Biggl[ 
{\rm tr} \Biggl\{ \hat{\cV}_i  
 \frac{\partial \hat{G}^\rR }{\partial \ve} \left( \hat{\cV}_j \hat{G}^\rR \hat{\cV}_k + \frac{1}{2} \hat{\cV}_{jk} \right) ( \hat{G}^\rR - \hat{G}^\rA ) \Biggr\}  
\nonumber \\
&\quad + (j \leftrightarrow k), 
\label{eq:NLC}
\end{align}
where $\hat{G}^\rR = (\ve - \hat{H}_\k - \hat{\Sigma}^\rR)^{-1} = (\hat{G}^\rA)^\dagger$ is retarded Green's function with self-energy $\hat{\Sigma}^{\rm R}$, and $\hat{\cV}_{ij} = \partial_{k_i} \partial_{k_j} \hat{H}_\k$. The trace runs over all of the orbital/band indices. For simplicity, we here consider the constant pure imaginary self-energy $\hat{\Sigma}^{\rm R} = -i/(2\tau)$. We here dropped the term whose integrand is proportional to $f(\ve)$ (not $df/d\ve$) because it vanishes in the time-reversal symmetric system~\cite{NYY2024}.

%%%%%%%%%%%%%%%%%%
\subsection{Symmetry argument}
\label{sec:symmetry}
%%%%%%%%%%%%%%%%%%

Before entering the quantitative discussion of the NCTE charge and thermal Hall effect and second-order response to the electric field, we first discuss the symmetry aspects of the B20-type crystals to discuss the qualitative feature of NCTE charge and thermal Hall effect. Table~\ref{tab:T} shows the character table of point group $T$, in which the space group $P2_1 3$ belongs. The irreducible representation $T$ contains the basis functions $(X,Y,Z)$ correspond to $(j_x,j_y,j_z)$, and quadratic functions correspond to the product of applied field(s). Therefore, this reveals the possible nonlinear responses as 
\begin{align}
(j_x, j_y, j_z) &= \sigma_1^{\rm E2} (E_y E_z, E_z E_x, E_x E_y) 
\\
%(j_x, j_y, j_z) &= 
&+
\sigma_1^{\rm ET} (E_y \partial_z T, E_z \partial_x T, E_x \partial_y T) \nonumber \\
&+
\sigma_2^{\rm ET} (E_z \partial_y T, E_x \partial_z T, E_y \partial_x T) .
\end{align}
Namely, $\s_{ijk}^{(2)}$ accepts only one independent component $\s_1^{\rm E2}$, and the response to the product of the electric field and the temperature gradient has two independent components, $\s_1^{\rm ET}$ and $\s_2^{\rm ET}$. The relation to the NCTE Hall conductivity is $\sigma_z^{\rm C} = (\s_1^{\rm ET} - \s_2^{\rm ET})/2$. The same discussion applies to the NCTE thermal Hall conductivity. 

Once we consider the current along [111] direction, the second-order response to the electric field vanishes, and the response to $E_i \partial_j T$ contains only antisymmetric part, $\propto {\bm E} \times \nabla T$, because of the three-fold rotational symmetry around [111] axis. This feature is similar to the chiral tellurium case we investigated previously~\cite{NYY2024}. Hence, we focus on the NCTE Hall currents along [111] direction in the calculations based on Wannier model. 

It is worth noting the relation between the NCTE Hall currents in the $z$ direction $j_z^{\rm C(T)}$ and the currents along the [111] direction $j_{[111]}^{\rm C(T)}$. The symmetric and anti-symmetric tensors do not change their symmetry after the coordinate transformation. Moreover, according to the symmetry arguments under $P2_1 3$ space group, we get the same NCTE Hall conductivities $\s_{[111]}^{\rm C(T)} = \s_z^{\rm C(T)}$. For instance, if we set $E_y = E_z = \partial_x T = \partial_z T = 0$, one can derive $j_{[111]}^{\rm C(T)} = \s_{[111]}^{\rm C(T)} ({\bm E} \times (-\nabla T)/T)_{[111]} = \sqrt{3} \s_z^{\rm C(T)} ({\bm E} \times (-\nabla T)/T)_z$, which is used for the numerical plot in the later discussion. 

One more interesting consequence of the symmetry argument is the absence of the Berry curvature dipole terms, Eqs.~\eqref{eq:NCTE_yC} and \eqref{eq:NCTE_yCT}. Since we could show that ${\bm v}_{n{\bm k}}= R_3^{-1} {\bm v}_{n} (R_3 {\bm k})$ and $\Omega_{n\bm k}^i = R_3^{-1} \Omega_{n}^i (R_3 {\bm k})$ where $R_3$ the three-fold rotational operation with respect to the [111] axis, we get $\Omega'_x = \Omega'_y = \Omega'_z$, leading to $\sigma_z^{\rm BCC} = 0$ and $\s_z^{\rm BCT} = 0$. Namely, both the equivalence of the $x$, $y$, and $z$ axes and the form of the Berry curvature dipole terms lead to the disappearance of $\s_z^{\rm BCC}$ and $\s_z^{\rm BCT}$ in the cubic chiral crystals. Therefore, the leading term of the NCTE charge and thermal Hall conductivities are solely governed by the orbital magnetic moment terms, 
\begin{align}
\s_z^{\rm C} \simeq \s_z^{\rm OMC},
\\
\s_z^{\rm T} \simeq \s_z^{\rm OMT}.
\end{align} 
This result can also be shown by numerical calculations. In our previous study, we obtained a similar conclusion for the isotropic minimal model~\cite{YNY2023} and in the vicinity of the top of the valence band in the chiral tellurium~\cite{NYY2024}, while this time, we confirmed this fact for a cubic chiral crystal. 

%%%%%%%%%%%%%%%%%%
\subsection{NCTE charge and thermal Hall effect}
\label{sec:NCTEC}
%%%%%%%%%%%%%%%%%%

In the Wannier model which faithfully reproduces the DFT bands, we quantitatively evaluate the NCTE charge and thermal Hall effects using Eqs.~\eqref{eq:NCTE_TOT}, \eqref{eq:NCTE_TOTT}, and \eqref{eq:NLC}. Figures \ref{fig:2}(c) and \ref{fig:2}(d) depict their chemical potential dependence of the NCTE charge and thermal Hall currents, respectively, along the [111] direction. We take the $k$-mesh number of $300^3$ for the momentum integral. The parameters are set as $E_x = 1100$~V/m, $k_{\rm B}T = 0.01$~eV and 0.03~eV, $\d_y T/ T = 100~{\rm m}^{-1}$, $\tau = 10$~{\rm fs}. Note that the symmetry consideration dictates the absence of second-order responses to electric fields in this direction. At $k_{\rm B}T=0.03$~eV, the magnitude of the charge Hall current is $\sim 0.27$~$\rm A/m^2$, and the thermal Hall current is $\sim 0.1$~$\rm W/m^2$ at the Fermi level, indicating sufficiently measurable values. Moreover, focusing on the chemical potential dependence reveals significant enhancements near the Fermi level hosting Rarita-Schwinger-Weyl (RSW) fermion and around $-0.2$~eV where double spin-1 excitation are located. These enhancements occur due to the dipole-like structures of orbital magnetic moments observed in Fig.~\ref{fig:3}, leading to $\nabla_\k \cdot {\bm m}_{n\k}^\perp \neq 0$. It is worth emphasizing once again that the three-fold rotational symmetry around the [111] axis precludes contributions from Berry curvature dipoles to the NCTE charge and thermal Hall effects in the B20-type compounds; leading term of the NCTE Hall currents (conductivities) are solely governed by the contribution from the orbital magnetic moment. Given that the NCTE Hall currents appearing in the [111] direction are the only Hall response in this direction, the NCTE Hall effect can be a very suitable transport measurement to discuss the effects of orbital magnetic moments arising from the multi-fold chiral fermions. The NCTE charge and thermal Hall effects are also enhanced due to the steep energy dependence of the density of states [Fig.~\ref{fig:2}(b)] as expected as the general property of the thermal transport; the NCTE charge Hall effect is interpreted as a first-order derivative and the NCTE thermal Hall effect behaves roughly as a second-order derivative of the density of states, as should be, because of the energy factors included in $F_{\rm C}(\ve_{n\k})$ and $F_{\rm T}(\ve_{n\k})$. 

Figure~\ref{fig:4} plots the temperature dependence of the NCTE charge Hall current $j_{[111]}^{\rm C}$ and thermal Hall current $j_{[111]}^{\rm T}$ at the Fermi level. One can clearly observe the sign change of $j_{[111]}^{\rm C}$ around the temperature $T \sim 0.01/k_{\rm B} = 116$~K which is not too low and relatively easy to access experimentally. The sign change occurs because of the odd function-like behavior in $\mu$ dependence of the NCTE charge Hall effect nearby the Fermi level, as shown in Fig.~\ref{fig:2}(c). We stress that this $\mu$ dependence is coming from the dipole-like distribution of orbital magnetic moment around the RSW fermion and other miscellaneous Weyl points reside in the vicinity of Fermi level. The similar behavior also appeared in the minimal model which exhibits NCTE charge Hall effect~\cite{YNY2023}. On the other hand, the even function-like behavior of $j_{[111]}^{\rm T}$ in the $\mu$ dependence around the Fermi level seen in Fig.~\ref{fig:2}(d) hinders any sign changes in the temperature dependence within the calculated range. This is natural because the $\mu$ dependence of $j_{[111]}^{\rm T}$ should be described as the energy derivative of $j_{[111]}^{\rm C}$, as pointed out in the previous paragraph. Meanwhile, the energy dependence of the NCTE charge (thermal) Hall effect is even (odd) function-like behavior around double spin-1 excitation located, making sign changes in the temperature dependence is absent (present); the tendency is opposite to the RSW fermion case. This also expresses the complexity of the momentum distribution of the orbital magnetic moment.

%%%%%%%%%%%%%%%%%%
\subsection{Second order response to the DC electric field}
\label{sec:NLC}
%%%%%%%%%%%%%%%%%%

Finally, we discuss the results of the second-order current response to the DC electric field. Figure~\ref{fig:5} plots the chemical potential dependence of each component of the normalized second-order nonlinear conductivity $\tilde{\sigma}_{ijk}^{(2)} \equiv \sigma_{ijk}^{(2)}/(e^3/h)$ at zero temperature. As expected from the symmetry arguments, the relationship $\sigma_{xyz}^{(2)} = \sigma_{yzx}^{(2)} = \sigma_{zxy}^{(2)} = \sigma_1^{\rm E2}$ is confirmed. Although small values remain for other components, these are artifacts resulting from slight symmetry breaking in the construction of the Wannier model. The estimated second-order current is $\sim 16.4~{\rm A/m^2}$, assuming the same value of applied DC electric field $E_x = E_y = E_z = 1100/\sqrt{2}~{\rm V/m}$, which is the same as the analysis of the NCTE (thermal) Hall effect. In magnetic B20-type compounds, conical magnetic structure causes nonreciprocal transport phenomena ($\sigma_{iii} \neq 0$)~\cite{Yokouchi2017,IN2020}, while in nonmagnetic (time-reversal symmetric) cases like CoSi, such longitudinal nonlinear transport is not expected. However, even without magnetism, the nonlinear Hall effects can persist finite due to the inversion symmetry breaking.

%%%%%%%%%%%%%%%%%%
\section{Summary}
\label{sec:summary}
%%%%%%%%%%%%%%%%%%

We investigated the nonlinear transport properties, particularly focusing on the NCTE charge and thermal Hall effects, in the B20-type compound CoSi. Based on band structures obtained from DFT calculations, we discussed the relationship between the NCTE Hall effects and the orbital magnetic moment. We found the distinctive behaviors of the orbital magnetic moment around topological singularities such as RSW fermions and double spin-1 excitations, significantly contributing to the NCTE charge and thermal Hall effects. Furthermore, we showed that the Berry curvature dipoles do not affect the NCTE Hall effects due to the crystal symmetry, highlighting the exclusive contribution of the orbital magnetic moment. We also revealed that the nonlinear current response along [111] direction is peculiar to the NCTE Hall responses and is absent for the second-order current response to the electric field. Additionally, we examined the temperature dependence of the NCTE charge and thermal Hall currents specifically around the Fermi level, and found the sign change in the NCTE charge and thermal Hall currents, that provide guidance for experimental studies. We unveiled that the NCTE Hall effects bring us the important information about the orbital magnetic moment emerging from the multi-fold chiral fermions in the cubic chiral crystals.  

\section*{acknowledgment}

The authors thank R. Iguchi, F. Kagawa, T. Nomoto, and K. Uchida for fruitful discussions. Parts of the numerical calculations have been done using the Supercomputer HOKUSAI BigWaterfall2 (HBW2), RIKEN. This work is supported by JSPS KAKENHI (Grant Nos.~JP20K03835, JP21K14526, and JP21K13875).

\end{document}